# Demonstration of TFTs 3D Monolithically Integrated on GaN HEMTs using Cascode Configuration with High Breakdown Voltage (>1900V)

Tian-Li Wu, *member IEEE*, Hsin-Jou Ho, Chia-Wei Liu, Yi-Chen Chen

*Abstract*— This study demonstrates 3D monolithic integration of amorphous indium-gallium-zinc oxide (a-IGZO) thin-film transistors (TFTs) on Gallium Nitride (GaN) high electron mobility transistors (HEMTs) in a cascode configuration, achieving high breakdown voltage capabilities exceeding 1900 V. Two device configurations, differing in a-IGZO channel thickness (30 nm / 10 nm), are fabricated and evaluated. Sample B, with a 10 nm a-IGZO channel, demonstrates superior electrical performance, including a high ON/OFF current ratio (~$10^7$), low subthreshold swing (SS), and a high breakdown voltage exceeding 1900 V comparable to standalone GaN power HEMTs. The results highlight the feasibility and potential of 3D integrated TFT on GaN power HEMTs, paving the way for new opportunities for the TFTs for high voltage applications.

## I. INTRODUCTION

The amorphous indium-gallium-zinc oxide (InGaZnO) thin film transistors (TFTs) have attracted great attentions due to their high mobility, transparency, and large-area uniformity by using a relatively low-temperature process [1][2], which can be fabricated by sputtering on arbitrary substrates for monolithically 3D integration. In addition, a-IGZO thin film features a wide bandgap of approximately 3.2 eV [3] that possesses high intrinsic breakdown voltage properties. Recently, high voltage a-IGZO TFTs have been demonstrated with promising performances, which are mainly attributed to the large extension of the gate-to-drain distances [4] and complicated designs of field plates to optimize the gate-to-drain electric field [5]. On the other hand, GaN high electron mobility transistors (HEMTs) also show significant potential for high voltage applications due to the wide bandgap and large electron density enabled by inherent two-dimensional electron gas (2DEG) [6][7]. Therefore, the combination of a-IGZO TFTs and GaN HEMTs can be a promising approach to take advantage of both technologies.

In this work, based on the superior properties with a-IGZO TFTs and GaN HEMTs, the high voltage TFT is demonstrated for the first time based on the monolithic 3D integration with GaN HEMTs to realize the cascode configuration. The cascode approach considers a low-voltage TFT with a high-voltage GaN HEMTs, where GaN HEMTs are mainly used to sustain high voltage. a-IGZO TFT is 3D monolithically integrated on the GaN power HEMTs to realize the cascode configuration, exhibiting more than >1900V breakdown voltage.

## II. DEVICE STRUCTURE AND FABRICATION

The cascode configuration has been widely adopted in recent years [6][7][8][9] in the package level to achieve a normally-off characteristic that combine a low voltage Si MOSFET with a high voltage GaN HEMT [6][8] or a low voltage GaN HEMT with a high voltage SiC JFETs [9]. **Fig. 1 (a)** shows the circuit diagrams for the cascode approach based on a low voltage TFT and high voltage GaN HEMTs and **Fig. 1(b)** shows the schematic of the low voltage TFT 3D monolithically integrated on GaN HEMTs to realize the cascode configuration [6]. During the ON-state when $V_{GS}$ is larger than $V_{TH,TFT}$, TFTs begin to turn on, creating a conductive path between the drain and source in TFTs. Furthermore, the HEMTs is also turned on due to the normally-on characteristics for HEMTs. During the OFF-state when $V_{GS}$ is smaller than $V_{TH,TFT}$ and a large $V_D$, the conductive channel is not formed in TFTs as well and the channel in HEMTs is depleted due to the negative $V_{DG}$. Therefore, HEMTs can sustain the high voltage due to the channel pinch-off.

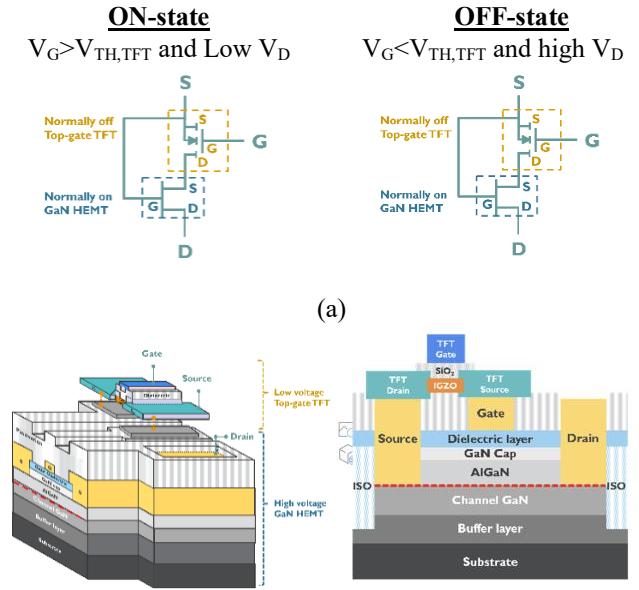

(a)

Manuscript received XXX, 2025; revised XXX, 2025; accepted XXX, 2025. Date of publication XXX, 2025; date of current version XXX, 2025. This work was financially supported by the "Advanced Semiconductor Technology Research Center" from The Featured Areas Research Center Program within the framework of the Higher Education Sprout Project by the Ministry of Education (MOE) in Taiwan. Also supported in part by the National Science and Technology Council (NSTC), Taiwan, under under Grant 112-2628-E-A49-020-MY3. The review of this letter was arranged by Editor XXX. (*Corresponding author: Tian-Li Wu,*)

T.-L. Wu is with Institute of Electronics, National Yang Ming Chiao Tung University, 30010 Hsinchu, Taiwan, and also with International College of Semiconductor Technology, National Yang Ming Chiao Tung University, 30010 Hsinchu, Taiwan. (e-mail: tlwu@nycu.edu.tw)

C.-W. Liu, H.-J. Hou, and Y.-C. Chen are with the International College of Semiconductor Technology, National Yang Ming Chiao Tung University, 30010 Hsinchu, Taiwan.

(b)

Fig. 1. Schematic of cascode configuration based on low voltage TFT and high voltage GaN HEMTs (a) and a-IGZO TFT directly stacks on GaN HEMTs to realize the cascode configuration (b).

The device fabrication process is divided into two parts: GaN HEMTs and a-IGZO TFTs, as shown in **Fig. 2**.

The HEMTs fabrication starts with device isolation by $N^+$ ion implantation. Next, a 30 nm SiON gate dielectric layer is deposited by PECVD. After that, ohmic contact areas are opened by an etching process and Ti/Al/Ti/Au (25/125/55/45 nm) is evaporated to serve as the source/drain metal for HEMTs. Rapid thermal annealing (RTA) is conducted at 850 °C to form the ohmic contacts. Finally, Ni/Au (20/100 nm) is evaporated to serve as the gate metal of HEMTs. The epitaxial layers consist of a 1000μm Si substrate, 4.3 μm buffer layer, 300 nm channel GaN, 26nm AlGaN, 1 nm GaN cap, and 2 nm *in-situ* SiN layer. After the HEMTs fabrication, a 200 nm $SiO_2$ layer is deposited as the passivation layer. Subsequently, a-IGZO channel with a thickness of 30nm (Sample A) and 10 nm (Sample B) (**Table 1**). are deposited by RF sputtering system, with a composition ratio of In:Ga:Zn:O = 1:1:1:4. The passivation layer on the contact area is then removed by a dry etching process, followed by Ti/Ni (25/25 nm) evaporation to form the source and drain of the IGZO TFTs. Next, a 50 nm $SiO_2$ gate dielectric and a 50 nm Ni gate metal are deposited, which also functions as the gate for the cascode structure. Finally, the process concludes with the passivation layer etching over the drain area of the HEMTs.

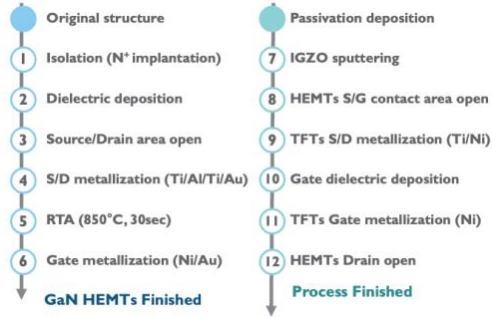

Fig. 2. Process flow of the TFTs monolithically 3D stacked on the GaN HEMTs for cascode configuration.

Table 1. Summary of the details fabricated in this work

|  | IGZO thickness |  |
| --- | --- | --- |
| Sample A | 30nm | GaN-on Si HEMTs |
| Sample B | 10nm | GaN-on-Si HEMTs |

### III. RESULTS AND DISCUSSION

**Fig. 3** and **Fig. 4** show the typical $I_D$-$V_G$ and $I_D$-$V_D$ characteristics in the fabricated TFTs 3D monolithically integrated on GaN HEMTs using cascode configuration. The well-behavior electrical characteristics can be observed. Sample A shows higher on-current but more negative $V_{TH}$ (-0.3V extracted at $I_D=10^{-5}$mA). On the other hand, Sample B shows a positive $V_{TH}$ (+0.53V extracted at $I_D=10^{-5}$mA) and better ON/OFF ratio (~$10^7$) compared to Sample A, which is mainly due to the thin TFT channel in Sample B. Furthermore, Sample B shows a similar $V_{TH}$ but with higher ON/OFF ratio compared to the typical TFT, which is mainly due to the low Ohmic contact resistance in the drain side connected in HEMTs and low leakage current in HEMTs.

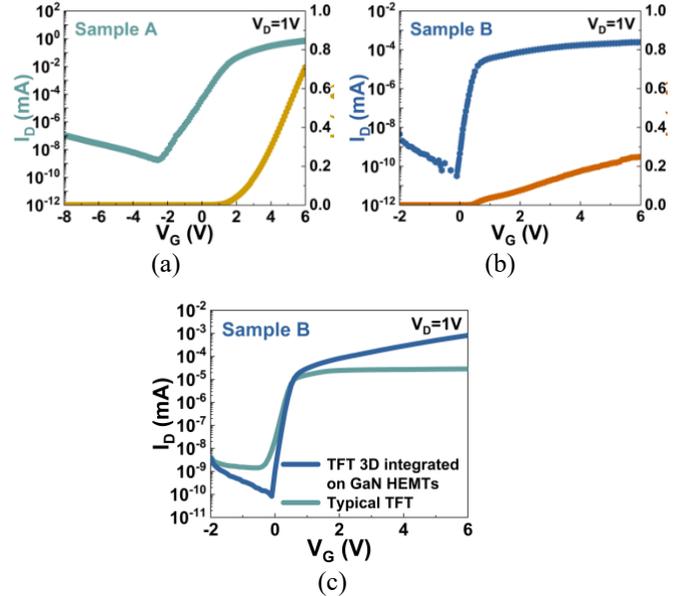

Fig. 3. $I_D$-$V_G$ characsersitics for Sample A (a) and Sample B (b) and the comparaion in Sample B with with a typical TFT (c).

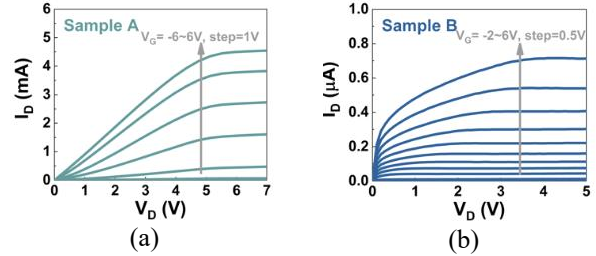

Fig. 4. $I_D$-$V_D$ charactersitics in Sample A (a) and Sample B (a).

**Fig. 5** shows the breakdown voltage charactestics in Sample B (a) and the compraions with the GaN power HEMTs (b) with same $L_{GD}$. Note that Sample A exhibits a high driain lekge current (not shown here) during the breakdown measurmeent at $V_G$=0V due to the depletion-mode charactersitics ($V_{TH}$<0V). Regaring to Sample B, with larger $L_{GD}$, the breakdown voltage can be increased beyond 1900V in the farbricated TFT using 3D intergrated cascode on GaN HEMTs (**Fig. 5(a)**). Besides, the breakdown votlage of the farbricated TFT using 3D intergrated cascode is compatable with the typical GaN power HEMTs (**Fig. 5(b)**)

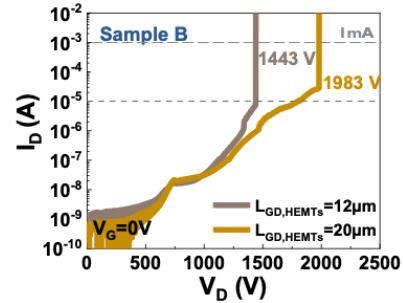

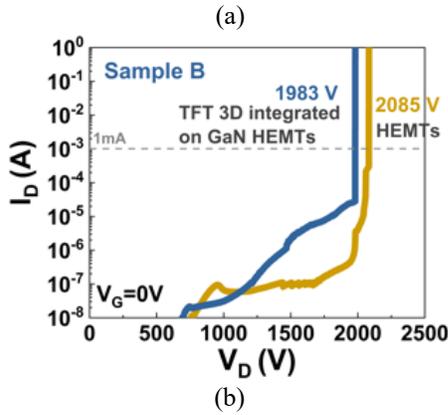

(a)

(b)

Fig. 5. Breakdown characteristics of (a) Sample B with $L_{GD}$ = 12 and 20 μm and comparison of the breakdown voltage with the typical power HEMTs (b).

**Table 2** shows the benchmark of this work compared other high voltage TFTs. The fabricated TFTs (Sample B) shows excellent SS, good On/Off ratio, and excellent breakdown voltage characteristics, indicating that 3D integrated TFT on GaN power HEMTs for the cascode configuration is a promising approach to boost the high voltage characteristics.

Table 2. Benchmark of high voltage TFTs compared with other works.

| Device Structures | Channel Material /Dielectric | Dimensions (μm) | $V_{TH}$ (V) | On Current | S.S. mV/dec | On/Off Ratio | Breakdown Voltage /Criteria | Ref. |
|---|---|---|---|---|---|---|---|---|
| TFT 3D integrated on GaN HEMTs (Sample B) | 10nm IGZO/ SiO$_2$ | W=200, L$_G$=20 | 0.58 | 8.1x10$^{-7}$ A (V$_D$=1V) | 90.3 | ~10$^7$ | 1983V/ 10$^{-3}$A | This works |
| GaN HEMTs | 2DEG/ SiON | W=100, L$_G$=5 | -4.47 | 46.38 mA/mm (V$_D$=1V) | 70.3 | ~10$^7$ | 2085V/ 10$^{-3}$A | This works |
| TFTs | 20nm IGZO with H$_2$ doped/SiO$_2$ | W=500, L$_G$=50, L$_{drift}$=5, L$_{dope}$=24 | / | ~10$^{-6}$ A (V$_D$=0.1V) | / | ~10$^7$ | 406V/1mA | [4] |
| TFTs w/ Field Plate | 50nm IGZO/SiO$_2$ | W=100, L$_G$=5 | / | ~10$^{-7}$ A (V$_D$=100V, V$_{FP}$=10V) | / | ~10$^6$ | 1902.5V /10uA /V$_{FP}$=-80V | [5] |
| TFTs | IGZO | W=60, L$_G$=200, L$_{Offset}$=100 | / | / | / | ~10$^5$ | 1100/12uA | [10] |
| TFTs | 10nm IGZO/Al$_2$O$_3$ | W=100, L$_{channel}$=5 L$_{GDoverlap}$=3 | 0.13 | 5 μA/μm (V$_D$=2V) | 190 | ~10$^9$ | 100/ 1A/μm | [11] |
| TFTs w/ stair Gate | 20nm IGZO/ stair-Al$_2$O$_3$ | W=200, L$_{channel}$=50 | 12.3 | ~10$^{-6}$ A (V$_D$=5V) | 129 | ~10$^7$ | 60/1mA | [12] |
| TFTs | 20nm IGZO with F doped/SiO$_2$ | W=20, L$_{GS}$=5, L$_{GD}$=1.5, L$_{channel}$=6 | / | ~10$^{-5}$ A (V$_{DS}$=1V) | / | ~10$^8$ | 91.6/1mA | [13] |

## IV. CONCLUSION

This work successfully integrates a-IGZO TFTs with GaN HEMTs using a 3D monolithic approach in a cascode configuration. The cascode configuration can leverages the a-IGZO TFTs to control the ON-state operation and the GaN HEMTs to sustain the high drain voltage during the OFF-state operation, achieving comparable breakdown voltage compared to the standalone GaN power HEMTs. The fabricated devices demonstrate promising electrical performance, with Sample B with a 10 nm a-IGZO channel showing a high ON/OFF ratio (~10$^7$), low subthreshold swing, and a breakdown voltage exceeding 1900 V. The superior performance of Sample B is attributed to 3D monolithic integration of TFTs on GaN-on-Si HEMTs, which can leverage the advantage of high breakdown capability of GaN technology. In sum, this study establishes the viability of 3D integration of TFTs and GaN HEMTs for high-voltage applications.